\documentstyle[aps,prl,preprint,floats,epsfig]{revtex}

\def\Dp{D^+} 

\def\ds{D_s^+}

\def\K0S{K^0_s}

\def\q2{q^2}
           
\def\r2{R_2}

\def\spin1{spin-one}

\def\gev2{{\rm GeV}^2}

\begin{document}

\preprint{\tighten\vbox{\hbox{\hfil CLNS 97/1515}
                        \hbox{\hfil CLEO 97-22}
}}

\title {\boldmath Measurement of the Branching Ratios for the Decays of $\ds$ to
 $\eta \pi^+,~\eta' \pi^+,~\eta \rho^+,$ and $\eta' \rho^+$}

\author{CLEO Collaboration}
\date{\today}

\maketitle
\tighten

\begin{abstract} 
Using a data sample with integrated 
luminosity of $\rm 3.9~fb^{-1}$
collected in $e^+ e^-$ annihilation with the CLEO-II detector 
at the Cornell
Electron Storage Ring, we have measured the 
branching ratios for the decay modes 
$\ds \to  (\eta, \eta') \pi^+$
and $\ds \to (\eta, \eta')  \rho^+$ relative to
$D_s^+ \to \phi \pi^+$.
These decay modes are among the 
most common hadronic decays of the $D_s^+,$
and can be related by factorization to the semileptonic decays
$\ds \to  (\eta,\eta') \ell^+ \nu_l$.
The results obtained are compared with previous CLEO
results and with the branching ratios 
measured for the related semileptonic decays.
We also report results on the Cabibbo-suppressed decays
of the $D^+$ to the same final states.

\end{abstract}
\newpage

{
\renewcommand{\thefootnote}{\fnsymbol{footnote}}

\begin{center}
C.~P.~Jessop,$^{1}$ K.~Lingel,$^{1}$ H.~Marsiske,$^{1}$
M.~L.~Perl,$^{1}$ S.~F.~Schaffner,$^{1}$ D.~Ugolini,$^{1}$
R.~Wang,$^{1}$ X.~Zhou,$^{1}$
T.~E.~Coan,$^{2}$ V.~Fadeyev,$^{2}$
% ...
\end{center}
 
\small
\begin{center}
$^{1}${Stanford Linear Accelerator Center, Stanford University, Stanford,
California 94309}\\
$^{2}${Southern Methodist University, Dallas, Texas 75275}\\
% ...
\end{center}

\setcounter{footnote}{0}
}
\newpage


\section{INTRODUCTION}

\begin{figure}  [hb]
\epsfxsize=9cm
\centerline{\epsffile{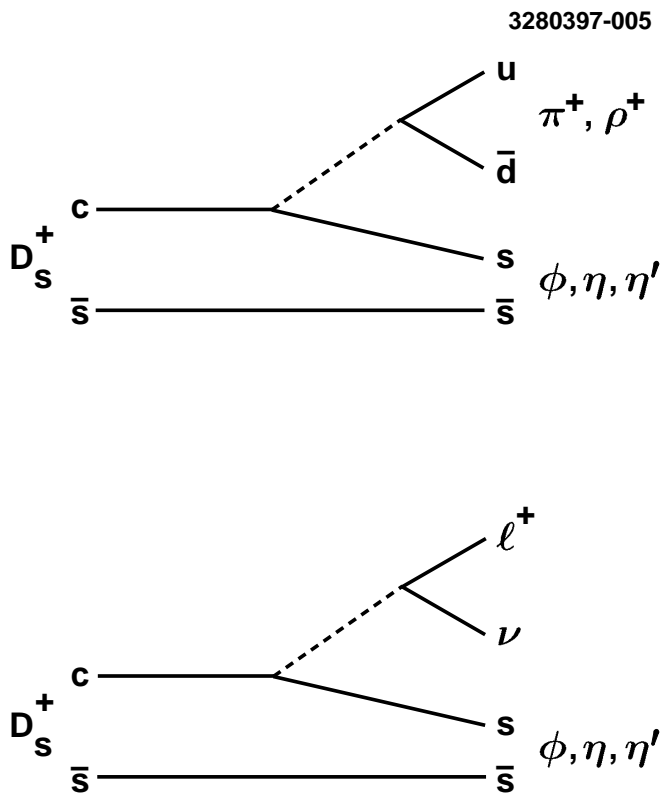}}
\vskip 1mm
\caption{Related Feynman digrams for hadronic and semileptonic $\ds$ decays}
\label{diags}
\end{figure}

Among the most common hadronic decay modes for the $\ds$
are the decays $\ds \to  (\eta, \eta') \pi^+$
and $\ds \to (\eta, \eta')  \rho^+$, where  
the notation $\ds \to  (\eta, \eta') \pi^+$
represents the decays $\ds \to  \eta \pi^+$
and $\ds \to  \eta' \pi^+$.  As can be seen from
Figure~\ref{diags},
they are related by factorization to the semileptonic decays
$\ds \to  (\eta,\eta') \ell^+ \nu_l$.  This relation
has been extensively discussed
by Kamal, Xu, and Czarnecki~\cite{Kamal1}.
One prediction of the factorization hypothesis
is that the $\ds$ decay
rate to  $\eta\rho^+$ can be simply related to the corresponding
semileptonic decay rate evaluated at $q^2 = m_{\rho}^2$:
\begin{equation}
	{\Gamma(\ds \to \eta \rho^+ )} = 6 \pi^2 a_1^2 f_{\rho}^2
|V_{ud}|^2~ {d\Gamma \over dq^2}(\ds \to \eta \ell^+ \nu_l) 
\vert_{ (q^2=m_{\rho}^2)}.
\label{eq4}
\end{equation}
Here $f_\rho$ is the decay constant for the $\rho$
and $a_1$ is a strong interaction coefficient that is
measured in two-body hadronic $D^0$ decays.

To test this factorization prediction experimentally, 
a shape for the form factor must be assumed.
It is expected to be very close to the form factor for
$D^0 \to K^- e^+\nu_e$, for which 
$${\Gamma(D^0 \to K^- e^+ \nu_e) \over
d\Gamma(D^0 \to K^- e^+ \nu_e) / dq^2}_{ (q^2=m_{\rho}^2)} = 1.30 \pm 0.01~\gev2.$$
This number is calculated using the CLEO measurement 
of the form factor~\cite{Klnu}.
Assuming a similar pole shape for the $\ds$ form factor yields
the prediction 
$\Gamma(\ds \to \eta \rho^+ )/\Gamma(\ds \to \eta e^+ \nu_e )  \approx 2.9$
and
$\Gamma(\ds \to \eta' \rho^+ )/\Gamma(\ds \to \eta' e^+ \nu_e)\approx 2.9$
~\cite{Kamal2}.

In 1992 CLEO~\cite{CLEOf} measured the branching ratios
for the hadronic modes studied here using a much smaller 
data sample of 0.69 ${\rm fb}^{-1}$.  
Combining these measurements with the more recent
CLEO measurements
of the semileptonic modes~\cite{CLEOl}, we calculate the 
$\Gamma(\ds \to \eta \rho^+ )/\Gamma(\ds \to \eta e^+ \nu_e )  
= 4.3 \pm 1.1$
and 
$\Gamma(\ds \to \eta' \rho^+ )/\Gamma(\ds \to \eta' e^+ \nu_e)
= 14.8 \pm 5.8$.
The last number is well above the factorization prediction of 2.9.
Theoretical efforts to understand
charm nonleptonic decays, even including final state
interactions, cannot account for the very large branching ratio
for $\ds \to \eta' \rho^+ $ ~\cite{HK,BLP}.

Because of the interest in these branching fractions,
we have remeasured them using the much larger data
sample now available.  In the present analysis we use data 
corresponding to an integrated luminosity of $\rm 3.9~fb^{-1}$ (which 
includes the 0.69 ${\rm fb}^{-1}$ used in the previous analysis) to remeasure
the four modes,
$\ds \to  (\eta, \eta') \pi^+$
and $\ds \to (\eta, \eta')  \rho^+$. 
The data were collected with the CLEO II detector at the Cornell
Electron Storage Ring (CESR), at center-of-mass energies equal to
the mass of the $\Upsilon$(4S) and in the continuum
just below the $\Upsilon$(4S) resonance.

The CLEO-II detector is designed to detect both charged and neutral
particles with high resolution and efficiency.  The detector
consists of a charged-particle tracking system surrounded by a
time-of-flight scintillator system.  These are in turn surrounded by
an electromagnetic calorimeter which consists of 7800 thallium-doped
CsI crystals.  This inner detector is immersed in a 1.5 T solenoidal
magnetic field generated by a superconducting coil.  Muon detection
is achieved using proportional tubes interleaved with iron.  A more
complete description of the detector can be found elsewhere~\cite{CLEOD}.

\section{EVENT SELECTION}
All events in this analysis are required to
pass standard CLEO criteria for hadronic events. 
Since all the signal modes involve only pions in the final
state, systematic errors are reduced by imposing no hadron identification 
cuts on either signal modes or the normalization mode.
All $\ds$ candidates are required to have 
$x = P_{D_s}/P_{max} > 0.63$ ($P_{max}^2 = E_{beam}^2 - M_{D_s}^2 $) 
to suppress
combinatoric background. Throughout this paper, 
reference to a particular charge state implies
the inclusion of the charge-conjugate state as well.

All photons are required to be in the good barrel 
region of the calorimeter ($|\cos\theta|<0.71$), to have a
minimum energy of 30 MeV,  and to not match the
projection of a charged track.  We choose pairs of photons whose
invariant mass is within 2.5 $\sigma(M)$ of the nominal
$\pi^0$ mass; $\sigma(M)$ is approximately 6~MeV/$c^2$.
We then kinematically constrain
the $\gamma \gamma$ pairs to the nominal 
$\pi^0$ mass in order to improve the momentum resolution of the
$\pi^0$ . We also require that $|\cos\theta_{\pi^0}|<0.8$,
where $\theta_{\pi^0}$ is the angle between one $\gamma$ in the 
$\pi^0$ rest frame and the $\pi^0$ momentum in the lab frame.
The signal is flat in $|\cos\theta_{\pi^0}|$ and the background
peaks toward +1.

For $\eta \to \gamma \gamma$ decays, the $\eta$ is 
selected in a manner similar to the $\pi^0$, but with the 
additional constraint that photons which could be paired to 
make $\pi^0$'s with momentum greater than 0.8~GeV/$c$ are rejected.
We also detect $\eta$'s using the $\eta \to \pi^+ \pi^- \pi^0$ 
decay chain, although this mode gives a sample with fewer events
and less significance than the two-photon decay mode.  
A $\pi^0$ momentum greater than 0.4~GeV/$c$ is required. 
All $\eta$ candidates within 2.5 $\sigma(M)$ of the nominal 
mass are considered,
where $\sigma(M)$ is the r.m.s.\ mass resolution for the 
given mode, typically about 14~MeV/$c^2$ for the 
$\gamma \gamma$ mode, and 6~MeV/$c^2$ for the $\pi^+ \pi^- \pi^0$ mode.
In order to improve the momentum resolution of the $\eta$, 
the decay particles from the $\eta$ are kinematically 
constrained to the nominal $\eta$ mass.

	To select $\eta'$ candidates we use the $\eta \pi^+ \pi^-$ final
state, where the $\eta$ is detected in both $\gamma \gamma $ and
$\pi^+ \pi^- \pi^0$ modes. Both $\eta$ and $\eta'$ candidates are
kinematically constrained to the nominal mass in order to improve
the momentum resolution.

Reconstruction efficiencies and invariant mass resolutions were
determined by using a GEANT-based ~\cite{GEANT} Monte Carlo (MC)
simulation of the detector.

\section {\boldmath $\ds$ Decays into modes containing a $\pi^+$ }

Five modes are studied in which a pion is produced in the weak decay:
\begin{enumerate}
\item $\ds \to \phi \pi^+$ (the normalization mode), $\phi \to K^+ K^-$
\item $\ds \to \eta \pi^+,~\eta \to \gamma \gamma $
\item $\ds \to \eta \pi^+,~\eta \to \pi^+ \pi^- \pi^0$
\item $\ds \to \eta' \pi^+,~\eta' \to \eta \pi^+ \pi^-,~\eta \to \gamma \gamma $
\item $\ds \to \eta' \pi^+,~\eta' \to \eta \pi^+ \pi^-,~\eta \to \pi^+ \pi^- 
\pi^0$
\end{enumerate}
We require the pions that come directly from the weak decay to
have momentum greater than
0.7~GeV/$c$ and the $\eta$ or $\eta'$ from the $\ds$ to have 
momentum greater than 1~GeV/$c$. This reduces the background from
random combinations. 

\subsection{\boldmath $\ds \to \phi \pi^+$}

\begin{figure}  [hbtf]
\epsfxsize=9cm
\centerline{\epsffile{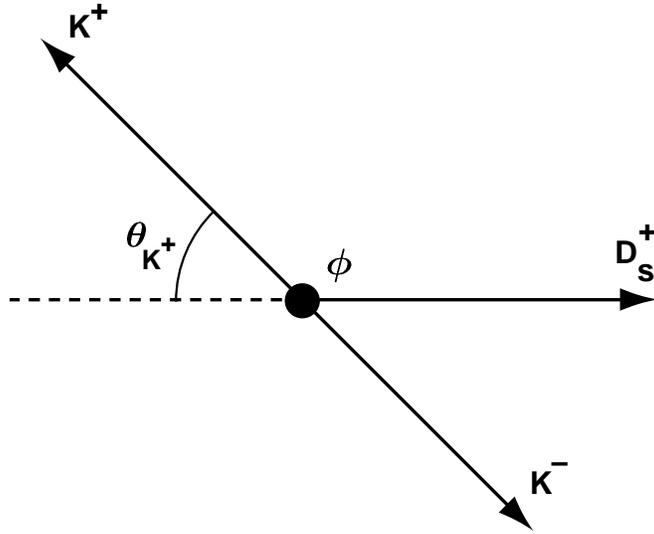}}
\vskip 1mm
\caption{ Illustration of the helicity angle, $\theta_{K^+}$. All
vectors represent momenta in the $\phi$ rest frame.}
\label{hel}
\end{figure}

Since this decay involves a pseudoscalar meson decaying 
into a vector meson and a pseudoscalar $\pi^+$, the $\phi$ 
must be polarized in the helicity zero state.
We take advantage of this by cutting on $\cos{\theta_{K^+}}$,
where $\theta_{K^+}$ is the angle between the $K^+$ momentum and 
the direction opposite to the $\ds$ momentum in the $\phi$ rest frame.
The angle is shown in Figure~\ref{hel}. 
The signal has a $\cos^2{\theta_{K^+}}$ distribution, 
while the background is flat in $\cos{\theta_{K^+}}$.
We require $|\cos\theta_{K^+}|>0.45$. 

\begin{figure}  [hbtf]
\epsfxsize=10cm
\centerline{\epsffile{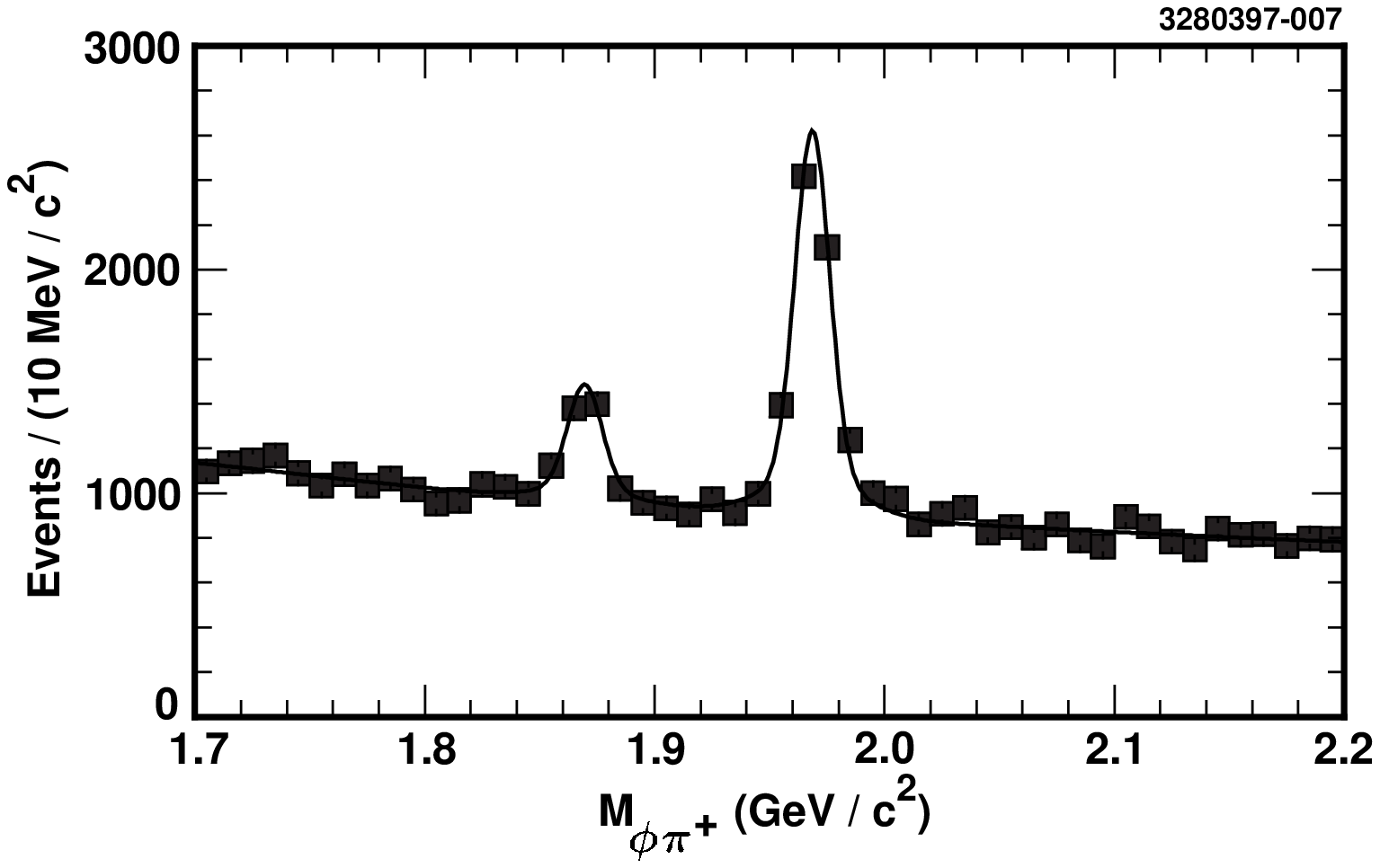}}
\vskip 1mm
\caption{ The $M(\phi \pi^+)$ distribution. 
The larger peak is due to the decay $\ds \to \phi \pi^+$; 
the smaller peak at lower mass is due to $\Dp \to \phi \pi^+$.}
\label{phipi}
\end{figure} 

We select $\phi$ mesons within $\pm$8 MeV of the peak mass, and
form the $\phi \pi^+$ mass spectrum shown in Figure~\ref{phipi}.
The $\phi \pi^+$ mass distribution 
shows two clear peaks, one from the $\ds$ and the
other from the $\Dp$.   To fit the spectrum
we use four functions:

\begin{enumerate}

\item The $\ds$ signal is fit to a sum of two Gaussians with a common mean; 
the widths and relative areas are fixed to values determined from the 
Monte-Carlo signal simulation.  
The mean is allowed to vary in the fit.

\item The $\Dp$ signal shape is of the same form as 
for the $\ds$, with the mass 
constrained to be 0.099~GeV/$c^2$ less than
the $\ds$ mass, which is the precisely measured mass 
difference ~\cite{PDG}.

\item The shape of the function used to represent 
the $\ds \to (\phi, \eta, \eta') \rho^+ $
feed-through is determined from Monte Carlo.
 This feed-through
causes a broad peak in the mass of the 
$(\phi, \eta, \eta') \pi^+ $ system
centered at 1.7~GeV/$c^2$, which is parameterized with a Gaussian. 
The normalization of the feed-through is determined from the
measurement of the branching ratio~\cite{CLEOf}. 

\item A second-order Chebyshev polynomial is used to represent the 
combinatoric background.

\end{enumerate}
This fit yields $3748\pm 91$ $\ds$ events.
In all other fits, four functions are also used, although
the combinatoric background shape depends on the particular mode.

\subsection{\boldmath $\ds \to \eta \pi^+$}

\begin{figure} [htb]
\epsfxsize=9cm
\centerline{\epsffile{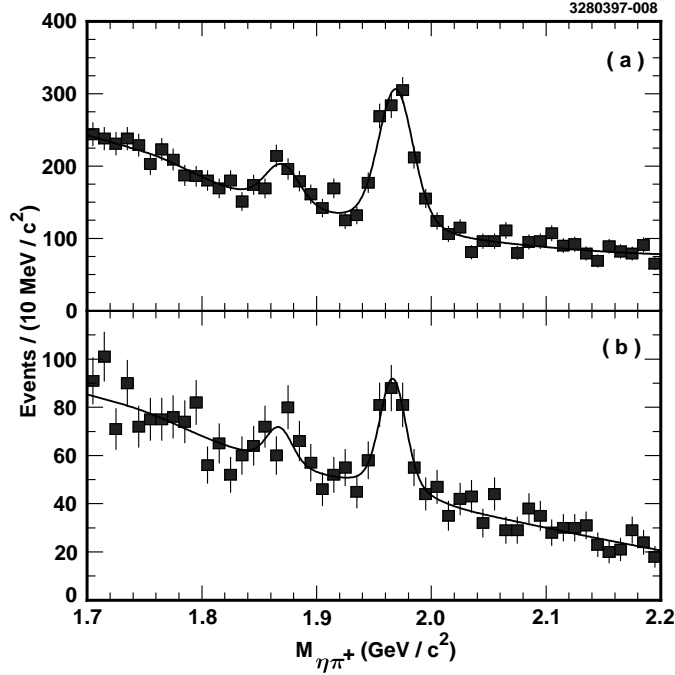}}
\vskip 1mm
\caption{The $M(\eta \pi^+)$ distribution for 
(a) $\eta \to \gamma \gamma$, (b) $\eta \to \pi^+ \pi^- \pi^0$. }
\label{etapi}
\end{figure}

	In Figure~\ref{etapi} we show the $\eta \pi^+$ invariant mass spectrum 
for both decay modes of the $\eta$.  The signal peaks are
evident for both the $\ds$ and $\Dp$. 
The peak at the $\ds$ contains $766\pm 44$ events for the 
channel $\eta \to \gamma \gamma$, and $154\pm 22$ events for
the channel  $\eta \to \pi^+ \pi^- \pi^0$. Multiple entries
into the plot from a single event are allowed, and no effort
is made to select among them.
In Table~\ref{tab:ds} we list the yields for
different channels and their efficiencies for $\ds$ decay.
We also list the measurement for
the ratio ${\Gamma(\ds \to \eta_{\gamma \gamma} \pi^+)/
\Gamma(\ds \to \phi \pi^+) }$.  In the table 
$\epsilon$ is the efficiency and
$\epsilon B$ is efficiency multiplied by
branching fraction of the secondary decays.
The systematic errors for the efficiencies relative to the $\phi\pi^+$ mode
have several sources and differ slightly from mode to mode. For the 
$\eta_{\gamma \gamma}\pi^+$ mode the systematic
error includes uncertainties in the relative charged track ($4\%$) 
and photon detection efficiencies ($5\%$). We studied the
Monte Carlo shape by letting the width of the two Gaussians vary
in the fit and then calculated the shift in the central value, giving us an
uncertainty of 3\%. We also used
different background shapes to determine the uncertainty due
to the unknown background shape, and obtained an error of $4\%$.
The total systematic error obtained by adding these
uncorrelated errors in quadrature is $8\%$.
For the $\eta_{3\pi}\pi^+$ mode the systematic
error includes uncertainties in the photon detection efficiency ($5\%$)
and in the signal ($5\%$) and 
background ($8\%$) shapes. In addition there was a systematic error
of ($10\%$) due to the modeling of multiple entries.
The total systematic error obtained by adding these
uncorrelated errors in quadrature is $15\%$.

\begin{table} [hbt]
\caption{Fit Results for $\ds \to  (\eta, \eta') \pi^+$. BR is the branching ratio of the $\phi (\eta, \eta')$ decay mode that is used. }
\label{tab:ds}
\begin{tabular}{cccccc}
Mode& N &$\epsilon$($\%$)&$\epsilon B$($\%$)&$\Gamma/\Gamma(\phi\pi)$
\\ \hline 
$\phi \pi$	   &$3748\pm 91$     & $19.1\pm 0.2$&9.4&\\
$\eta_{\gamma\gamma}\pi$&$766 \pm 44$&$9.6\pm 0.1$ &3.7&$0.52\pm 0.03\pm 0.04$\\
$\eta_{3\pi}\pi$&$154 \pm 22$        & $4.5\pm 0.1$ &1.1&$0.35\pm 0.05\pm 0.06$\\
$\eta'(\eta_{\gamma\gamma})\pi$&$479\pm 26$&$6.7\pm 0.1$&1.1&$1.09\pm 0.06\pm0.07$\\
$\eta'(\eta_{3\pi})\pi$&$58\pm 9$          &$1.9\pm 0.1$&0.2&$0.73\pm 0.11\pm 0.12$\\ \hline
\end{tabular}
\end{table}

\begin{table} [hbt]
\caption{Fit Results for $\Dp \to  (\eta, \eta') \pi^+$.}
\label{tab:dp}
\begin{tabular}{cccccc}
Mode & N&$\epsilon$($\%$)&$\epsilon B$($\%$)&$\Gamma/\Gamma(\phi\pi)$
\\ \hline
$\phi \pi$ &$1133\pm 72$     & $20.3\pm 0.2$&9.9&\\
$\eta_{\gamma\gamma}\pi$&$225 \pm 38$&$9.6\pm 0.2$ &3.7&
$0.53\pm 0.09\pm 0.05$\\
$\eta_{3\pi}\pi$&$50 \pm 20$ &$4.6\pm 0.1$ &1.1&$0.40\pm 0.15\pm 0.07$\\
$\eta'(\eta_{\gamma\gamma})\pi$&$114\pm 18$&$6.8\pm 0.1$&1.1&
$0.90\pm 0.14\pm 0.07$\\
$\eta'(\eta_{3\pi})\pi$&$12\pm7$ &$1.9\pm 0.1$&0.2&$0.52\pm 0.29\pm 0.09$\\ \hline
\end{tabular}
\end{table}

The measured ratio for ${\Gamma(\ds \to \eta_{3\pi} \pi^+)/
\Gamma(\ds \to \phi \pi^+) }$ shown in Table~\ref{tab:ds} is 
approximately two standard deviations lower than  
the corresponding ratio for the $\eta_{\gamma \gamma}$ mode,
taking into account the systematic errors which are not in
common to the two modes.  The measurements of $\ds \to \eta' \pi^+$, 
described in the next section, 
show a similar discrepancy, as do 
the $\Dp \to \eta \pi^+$ and $\eta' \pi^+$, although
those have less statistical significance.  
As a result, we searched in some detail for a systematic 
discrepancy in reconstructing the two 
$\eta$ decay modes.
To calibrate the relative efficiency for these modes, 
and to check the reconstruction program, we studied 
events of the type
$D^{*+} \to D^0 \pi^+ $ with $D^0 \to  \overline{K}^{*0} \eta $.
This has a very large and clean signal, and an $\eta$ momentum spectrum
very similar to that for the $\ds$ decays.
Using this process, we measure 
$B(\eta \to \gamma \gamma) /
B(\eta \to \pi^+ \pi^- \pi^0) = 1.53 \pm 0.16 \pm 0.10$,
compared to the PDG value of $1.64 \pm 0.04$~\cite{PDG}.
This confirms that the relative efficiency for the two decay modes
of the $\eta$ is reproduced properly in the Monte Carlo simulation.
Other checks using the data also reproduced the expected ratio
of $B(\eta \to \gamma \gamma) / B(\eta \to \pi^+ \pi^- \pi^0)$,
although with limited statistical power.  Since we were unable
to isolate any systematic effect, 
we attribute the difference between the two
$\ds \to \eta \pi^+$ measurements to an unlikely set
of statistical fluctuations.

The yields and relative branching ratios for all of the
$D^+$ decays into the same final states are shown in Table~\ref{tab:dp}.
The efficiencies for the $D^+$ modes
are generally very close to those for the corresponding $\ds$
decays.

\subsection{\boldmath $\ds \to \eta' \pi^+$}

For this mode, we can apply cuts on both 
the $\eta$ mass and the $\eta'$ mass, reducing the
background substantially.  Each mass provides
a kinematic constraint, helping to improve the resolution
for the $\eta' \pi^+$ mass.  As a result, these modes
are significantly cleaner than $\ds \to \eta \pi^+$.
We require the momentum of the $\eta'$  to be
greater than 1.0 GeV/c. 

\begin{figure}  [htb]
\epsfxsize=10cm
\centerline{\epsffile{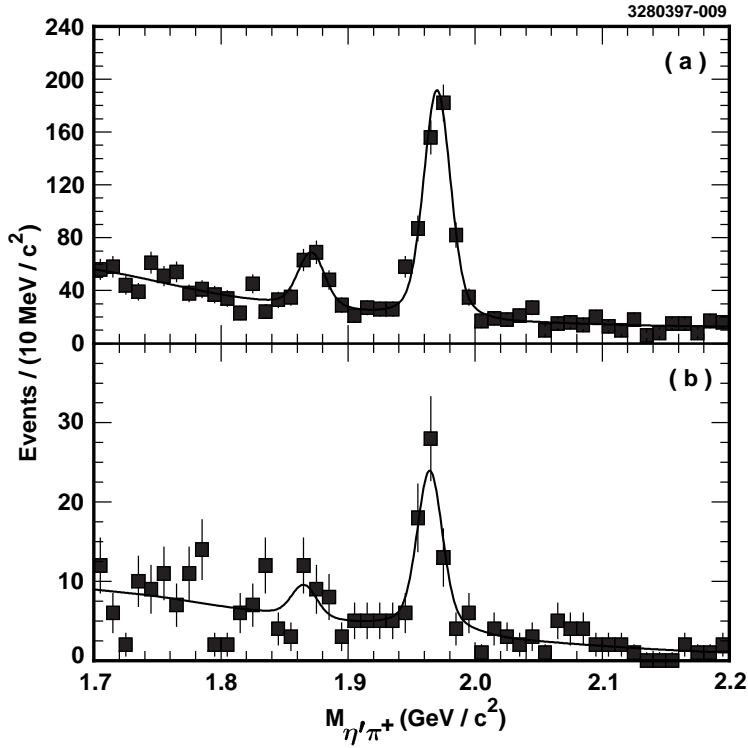}}
\vskip 1mm
\caption{The $M(\eta' \pi^+)$ distribution, using the decay mode $\eta'
\to \eta \pi^+ \pi^-$ with (a) $\eta \to \gamma \gamma$, (b) $\eta \to \pi^+ \pi^- \pi^0$.}
\label{etappi}
\end{figure}

In Figure~\ref{etappi} we show the $\eta' \pi^+$ invariant mass spectrum 
for both $\eta$ decay modes. 
The peak at the $\ds$ mass contains $479\pm 26$ events for the 
channel $\eta \to \gamma \gamma$, and $58\pm 9$ events for
the channel  $\eta \to \pi^+ \pi^- \pi^0$.
The efficiencies and relative branching ratios are shown
in Table~\ref{tab:ds}.
The systematic error on the branching ratio
measurement due to the uncertainty in charged track efficiency 
is negligible for the case of $\eta \to \gamma \gamma$ because
the final state has the same number of charged tracks as the normalizing mode.
The main contributions to the systematic error
are the uncertainties in the photon detection efficiency ($5\%$) and 
in the shapes used to describe signal ($3\%$) and background ($3\%$).
The total systematic error obtained by adding these
uncorrelated errors in quadrature 
is $6\%$. For the channel $\eta \to \pi^+ \pi^- \pi^0$,
the main contributions to the systematic error are
the uncertainties in the efficiency for charged tracks ($4\%$)
and photons ($5\%$) and 
in the shapes for signal ($10\%$) and
background ($4\%$), and in 
handling of multiple entries ($10\%$).
The total systematic error obtained by adding these
uncorrelated errors in quadrature is $16\%$.

        In ${D_s}^+ \to \eta' \pi^+$, $\eta' \to \eta \pi^+ \pi^-$, 
$\eta \to \pi^+ \pi^- \pi^0$, we found that there are many events
with multiple combinations of pions which satisfy our selection
criteria. Most of them come from real $\eta'$ decays in which 
different rearrangements of the same four charged pions 
(two directly from the $\eta'$ and two from the $\eta$), plus 
the $\pi^0$, pass our $\eta$ and $\eta'$ cuts. 
We take only one candidate 
per event, choosing the candidate with
the minimum value of a $\chi^2$ based on the $\pi^0$, $\eta$, and $\eta'$ 
masses: $\chi^2 = {(\delta M_{\eta'})^2 \over \sigma_{\eta'}^2} + 
{(\delta M_{\eta})^2 \over \sigma_{\eta}^2} +
{(\delta M_{\pi^0})^2 \over \sigma_{\pi^0}^2 }$.
The resulting measurements are shown in Tables~\ref{tab:ds}
and \ref{tab:dp}.

\section{\boldmath $\ds$ Decays into modes containing a $\rho^+$ }

The analogous $\ds$ decay channels, where the $\pi^+$ has been 
replaced by a $\rho^+$, can be studied using very similar cuts.
Because of lower rates, lower efficiency, and a serious
problem with multiple combinations within the same event,
the $\eta \to \pi^+\pi^-\pi^0$ decay does not add
significantly to the measurements of these modes,
and is not used.  A data sample with 
about $20\%$ less integrated luminosity was used for the measurements
of these modes.
\subsection{\boldmath $\ds \to \eta \rho^+$}

\begin{figure}  [htb]
\epsfxsize=10cm
\centerline{\epsffile{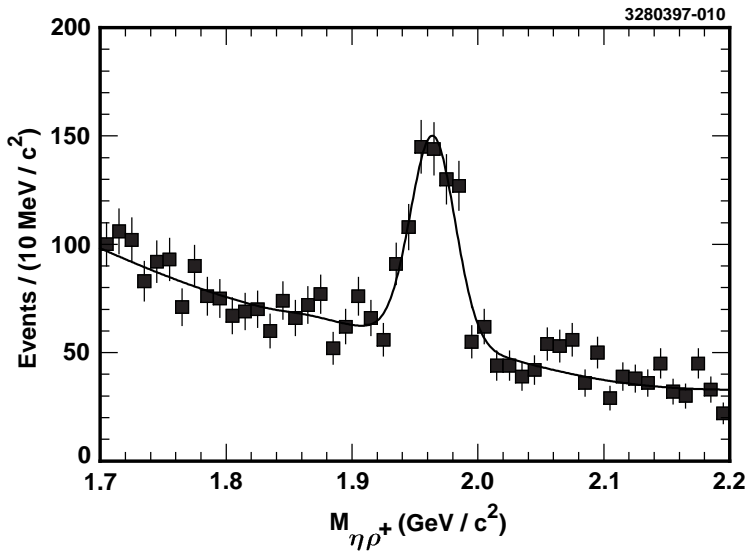}}
\vskip 1mm
\caption{The $M(\eta \rho^+)$ distribution, with 
$\eta \to \gamma \gamma$.}
\label{etarho}
\end{figure}

For the decay mode $\ds \to \eta \rho^+$,
we need to consider the possibility
of nonresonant $\eta \pi^+ \pi^0$ feedthrough.  
For Figure~\ref{etarho},
we require the helicity angle to be in the 
range \hbox{$|\cos\theta_{\pi^+}|>0.45$},
and the invariant mass of the $\pi^+ \pi^0$ 
to be within $\pm$170 MeV/$c^2$ of the $\rho^+$ mass.
A fit to the resulting $ \eta \pi^+ \pi^0$ mass 
spectrum is shown,
yielding $589\pm 43$ $\ds \to \eta \rho^+$ candidates
and $8\pm 32$ $\Dp \to \eta \rho^+$ candidates;
thus there is no evidence of $\Dp \to \eta \rho^+$.
We cannot directly extract a branching ratio for
$\ds \to \eta \rho^+$, however, until we account for
possible nonresonant feedthrough.

\begin{figure}  
\epsfxsize=16cm
\centerline{\epsffile{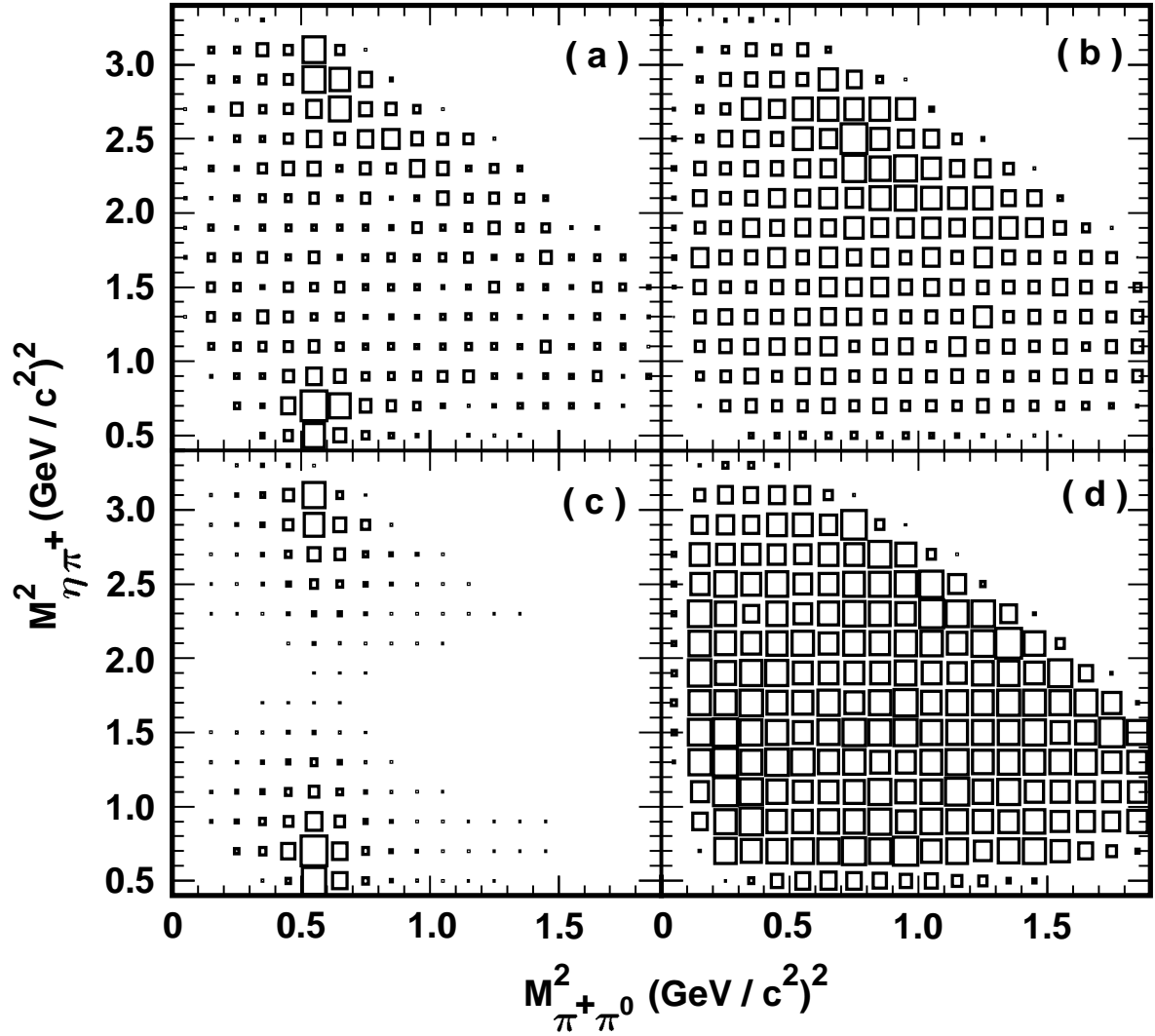}}
\vskip 1mm
\caption{Dalitz plot of $\ds \to \eta \rho^+$, with $\eta \to \gamma \gamma$.
The horizontal axis is $M_{\pi^+\pi^0}^2$; the vertical axis is
$M_{\eta \pi^+}^2$.
(a) Data signal region; (b) Data $M_{D_s}$ sidebands; 
(c) MC signal;
(d) MC simulation of nonresonant $ \eta \pi^+ \pi^0$, generated
according to phase space.}
\label{dalitz}
\end{figure}

Although cuts on the helicity angle and on
the $\rho$ mass region can be used, the most reliable way to measure
the resonant branching ratio is to fit the Dalitz plot.  
By doing this we make full use of the di-pion mass
and the helicity angle to isolate the $\eta \rho^+$ signal.
 We therefore make a Dalitz plot of all events with 
$ 1.94< M(\eta \pi^+ \pi^0)< 1.99$~GeV/$c^2$, removing 
the cuts on helicity angle and
on the $\pi^+\pi^0$ mass.
In Figure~\ref{dalitz} we show four Dalitz plots:
(a) the signal region in the data,
defined as $ 1.94< M(\eta \pi^+ \pi^0)< 1.99$~GeV/$c^2$;
(b) the $M_{D_s}$ data sidebands, which are the mass regions
	$1.75< M(\eta \pi^+ \pi^0)< 1.90$~GeV/$c^2$ and
	$2.04< M(\eta \pi^+ \pi^0)< 2.24$~GeV/$c^2$;
(c) the full Monte Carlo of the $ \eta \rho^+$ signal; and 
(d) a simulation using a parameterized Monte Carlo 
of nonresonant
$ \eta \pi^+ \pi^0$ events generated according to phase space.

We do not expect $\eta \pi$ resonant structures in this Dalitz
plot because isospin forbids $s\bar{s} \to \eta \pi$.
For all four Dalitz plots, we recalculate the values of 
$M^2(\eta \pi^+)$ and $M^2(\pi^+ \pi^0)$ so that the 
Dalitz boundary corresponds exactly to that of the mass of 
the $\ds$ ~\cite{PDG}, giving the sidebands the same
boundary as the signal region.  This causes negligible
smearing of the $\rho^+$ resonance.

The most obvious feature of the Dalitz plot is that the 
$\rho^+$ region stands out so clearly in the data, 
even though there is a significant non-$D_s^+$ background which contains
very little $\rho^+$.
A binned Dalitz fit to the data distribution 
in the signal region was performed using the 
sum of the distributions in the other three plots in 
Figure~\ref{dalitz}.  The normalization of the non-$D_s^+$ component
is fixed using a fit to the $\eta \rho^+$ mass distribution as in 
Figure~\ref{etarho} but without helicity angle and $\rho$ mass
cuts.
The number of the resonant and nonresonant $D_s^+$ events
is varied in the fit, with no interference term allowed.
The results of the fit are shown in Table~\ref{tab3}.
The systematic error includes
uncertainties in the efficiencies for charged tracks ($4\%$)
and photons ($10\%$) and the shapes for the 
signal ($4\%$) and the background ($3\%$).

In order to understand the systematic error due to
possible interference between
the resonant and nonresonant decays we also did
a coherent Dalitz fit.  The density of the events
in the Dalitz plot is represented by the expression
$$I=A_1^2 +A_2^2 + B\times 2A_1A_2\cos(\delta_1 - \delta_2),$$
where $A_1$ and $\delta_1$ are the amplitude and 
phase of the Breit-Wigner 
resonance, $A_2$ and $\delta_2$ are the amplitude and phase of the
nonresonant decay, both of which are assumed to be constant,
and $B$ is an additional constant which is allowed to vary from 
zero to one. 
The case $B=0$ corresponds to no interference
between the resonant and the nonresonant part; the
case $B=1$ corresponds to full interference,
expected if the nonresonant case were indeed
a single partial wave with constant phase. 
The true case could lie anywhere between these two limits.
In the fit when the constant $B$ is allowed to
float it takes the value $0.24\pm 0.20$, consistent
with no interference.  We therefore use the result
from the incoherent fit to determine the branching ratio,
and use the result from the coherent fit with $B=0.44$
to find a conservative systematic error from this source.
This corresponds to a $3.6\%$ error.
The total systematic error, obtained by adding this error 
in quadrature with the other systematic errors mentioned above,
is estimated to be $12\%$.

\subsection{\boldmath $\ds \to \eta' \rho^+$}

\begin{figure}  
\epsfxsize=10cm
\centerline{\epsffile{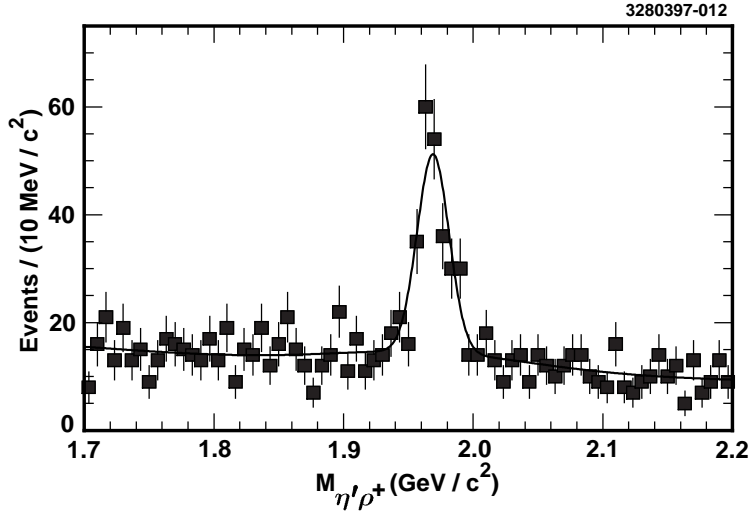}}
\vskip 1mm
\caption{The $M(\eta'\rho^+)$ distribution, with $\eta'
\to \eta \pi^+ \pi^-$, $\eta \to \gamma \gamma$.}
\label{etaprho}
\end{figure}

The decay $\ds \to \eta' \rho^+$ was reconstructed using the decay mode
$\eta' \to \eta \pi^+ \pi^-$, with $\eta \to \gamma \gamma$. 
We require the momentum of the $\eta'$ to be greater than 1.0~GeV/$c$ 
and the invariant mass of the two pions to be within $\pm$170 MeV of 
the $\rho^+$ mass.
In Figure~\ref{etaprho} we can see a clear peak of $M_{\eta' \pi^+ \pi^0}$.
The fit yields $181\pm 18$ $\ds \to \eta' \rho^+$ events and
$-4\pm 10$ $\Dp \to \eta' \rho^+$ events; thus there is no evidence for
$\Dp \to \eta' \rho^+$.

As for the case of the $\eta \rho^+$ decay mode, 
we need to subtract any nonresonant feedthrough into
the $\eta'\rho^+$ final state.  In this case, however,
a Dalitz plot is not as useful 
in separating signal from background, because
the kinematic range for the di-pion mass does not extend beyond
the region of the $\rho$.  
We do not expect $\eta' \pi$ resonant structures in this Dalitz
plot because isospin forbids $s\bar{s} \to \eta' \pi$. 
We therefore fit 
the angular distribution alone to extract the $\rho$ component. 
As for the case of the Dalitz fit for $\eta \rho^+$, we use
three components in the fit:
(a) the resonant signal shape, a fourth-order polynomial
determined from the Monte Carlo, which includes the distortion of
the pure $\cos^2\theta_{\pi^+}$ shape due to detector acceptance;
(b) a nonresonant $D_s^+$ shape, which is linear; and
(c) a non-$D_s^+$ background shape, which is a first-order polynomial 
determined by fitting the sidebands.
As in the Dalitz fit, we fix the background normalization from
the $\ds$ mass fit, and vary the normalizations  of
the signal and nonresonant parts. 

\begin{figure}  
\epsfxsize=10cm
\centerline{\epsffile{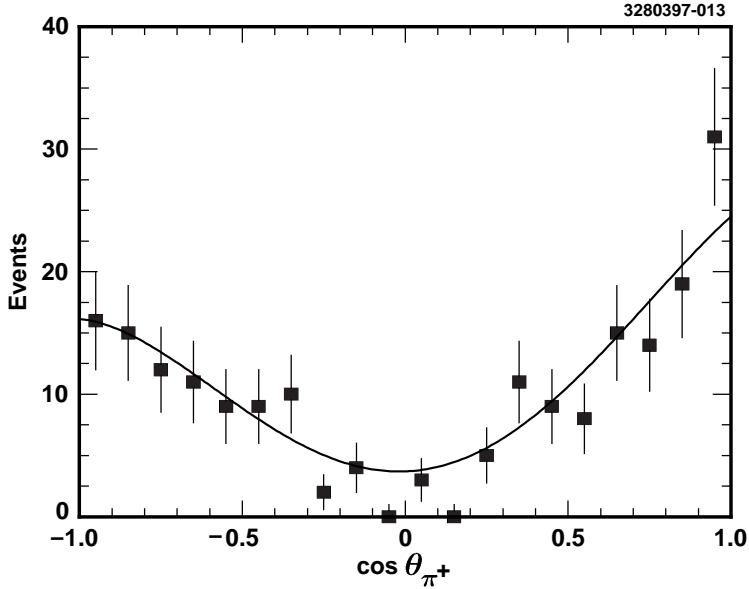}}
\vskip 1mm
\caption{The helicity angle distribution for events in the $\ds$ mass 
peak for the decay channel $\ds \to \eta' \rho^+$.}
\label{cos}
\end{figure}

Figure~\ref{cos} shows the fit of 
the helicity angle distribution for the events in the $\ds$ mass peak.
The results of the fit are shown in 
Table~\ref{tab3}.
The total systematic error  of $11\%$ includes uncertainties in the 
photon detection 
efficiency ($10\%$) and
in the signal ($4\%$), background ($3\%$), 
and nonresonant ($2\%$) shapes.

\bigskip
\vskip 4mm
\begin{table} [htb]
\caption{Fit Results for $\ds \to (\eta, \eta')  \rho^+$. }
\label{tab3}
\begin{tabular}{cccccc}
Mode & N&$\epsilon$($\%$)&$\epsilon B$($\%$)&$\Gamma/\Gamma(\phi\pi)$
\\ \hline 
$\phi\pi^+ $		    &$3000\pm81$&$19.1\pm0.2$&9.4&        \\
$\eta \rho^+$ & $447\pm 31$&$1.82\pm 0.07$ &0.47&$2.98\pm 0.20\pm 0.39$\\
$\eta'\rho^+$ & $137\pm 14$&$1.36\pm 0.04$&0.15&$2.78\pm 0.28\pm 0.30$\\\hline
\end{tabular}
\end{table}

The yields and upper limits on the branching ratios for the
$D^+$ decays into final states with a $\rho^+$ 
are shown in Table~\ref{tab4}.

\begin{table} [htb]
\caption{Fit Results for $\Dp \to (\eta, \eta')  \rho^+$. }
\label{tab4}
\begin{tabular}{cccccc}
Type& N&$\epsilon$($\%$)&$\epsilon B$($\%$)&$\Gamma/\Gamma(\phi\pi)(90\% C.L)$ \\ \hline 
$\phi\pi^+ $    &$970\pm65$&$20.3\pm0.2$&9.9&        \\
$\eta \rho^+$  & $8\pm 32$&$2.1\pm 0.1$ &0.55&$<$1.11 \\
$\eta'\rho^+$  & $-4\pm10$&$1.7\pm 0.1$&0.19&$<$0.86 \\ \hline
\end{tabular}
\end{table}

\section{Conclusions}
We have measured with improved statistics the branching ratios
of the two-body hadronic decays of the $\ds$:
$\ds \to  \eta\pi^+ ,~\eta'\pi^+,~\eta \rho^+,~{\rm and }~  \eta' \rho^+$. 
The results are consistent
with the previous CLEO measurements~\cite{CLEOf} and have improved errors. 
Using weighted averages of the two $\eta$ modes, our results for
$\ds \to (\eta, \eta') \pi^+$ are:
$${\Gamma(\ds \to \eta \pi^+)\over
\Gamma(\ds \to \phi \pi^+) }= 0.48\pm 0.03 \pm 0.04$$
and 
$${\Gamma(\ds \to \eta' \pi^+)\over
\Gamma(\ds \to \phi \pi^+) }= 1.03\pm 0.06 \pm 0.07.$$
The results for the $\rho$ modes are:
$${\Gamma(\ds \to \eta \rho^+)\over
\Gamma(\ds \to \phi \pi^+) }= 2.98\pm 0.20 \pm 0.39$$
and
$${\Gamma(\ds \to \eta' \rho^+)\over
\Gamma(\ds \to \phi \pi^+) }= 2.78\pm 0.28 \pm 0.30.$$

Using these measurements and the published CLEO semileptonic
measurements~\cite{CLEOl}, we can calculate the ratios 
which test factorization:  
$\Gamma(\ds \to \eta \rho^+ )/\Gamma(\ds \to \eta e^+ \nu_e )  
= 4.4 \pm 1.2$
and 
$\Gamma(\ds \to \eta' \rho^+ )/\Gamma(\ds \to \eta' e^+ \nu_e)
= 12.0 \pm 4.3$.
The branching ratio for the mode
$\ds \to \eta' \rho^+$ is much larger than the value of 2.9
expected from factorization.  
Using the normalization 
$B(\ds \to \phi \pi^+)= (3.6 \pm 0.9)\%$~\cite{PDG}, we calculate
${B(\ds \to \eta' \rho^+)}= (10.0 \pm 1.5\pm 2.5)\%$, where the
second error is due to the uncertainty in the $\ds \to \phi \pi^+$
branching fraction.
This branching fraction is very large, considering the
flavor wave function of the $\eta'$ is only partly $s \bar s$
and that the rate is suppressed for such 
a $P$-wave decay very close to threshold.
There is no obvious mechanism by which final state interactions
could cause such a large enhancement of one of the dominant decay modes.

Table~\ref{ds_comp}
summarizes the measurements of
branching ratios for all four $\ds$ decays and compares them
with theoretical calculations.
Models which are successful in predicting other charm hadronic modes
reasonably well predict ${B(\ds \to \eta' \rho^+)}$
to be $1-3\%$ ~\cite{HK,BLP,BSW}.
This failure leads theorists to consider contributions
to the amplitude from decay
diagrams other than that shown in Figure 1. 
For example, Ball {\it et al.} ~\cite{Ball} argue that the high branching
ratio for $\ds \to \eta' \rho$ could be due to
a $c \bar s$  annihilation into a $W^+$ and two gluons,
in which the two gluons hadronize as an $\eta'$.

\begin{table} [hbt]
\caption{Measurements and Predictions for Branching Ratios of $\ds$ Decays.}
\label{ds_comp}
\begin{tabular}{ccccc}

Mode&$\Gamma/\Gamma(\phi\pi^+)$& BSW~\cite{BSW}&HK~\cite{HK}&BLP~\cite{BLP}\\ \hline
$\eta \pi$ & $0.48\pm0.05$     &1.04&$0.58\pm0.15$&0.30\\
$\eta \rho$& $2.98\pm0.44$     &1.96&$2.86\pm0.71$&1.83\\
$\eta' \pi$& $1.03\pm0.09$     &0.61&$1.55\pm0.42$&1.32\\
$\eta'\rho$&$2.78\pm0.41$      &0.55&$0.43^{+0.55}_{-0.32}$&0.59\\ \hline
\end{tabular}
\end{table}

Using the normalization $B(\Dp \to \phi \pi^+) = 
(6.1\pm 0.6) \times 10^{-3}$ ~\cite{PDG},
we also calculate the $D^+$ branching fractions to the same final states.
Table~\ref{dp_comp} summarizes the results. 
Since the $\Dp$ decays involve two diagrams which interfere,
the theoretical calculations vary widely, and are expected to be
somewhat less reliable than for the $\ds$ case.
\begin{table} [hbt]
\caption{Measurements and Predictions for Branching Fractions of $D^+$
Decays. The experimental upper limits are at the $90\%$ confidence level.}
\label{dp_comp}
\begin{tabular}{ccccc}

Mode &Branching Fraction (\%) &BSW~\cite{BSW}&HK~\cite{HK}&BLP~\cite{BLP}\\ \hline
$\eta \pi$ &$0.30\pm 0.06$&0.004&$0.68\pm0.21$&0.34\\
$\eta \rho$&$<0.68 $          &0.06 &$<0.48$ & 0.01\\
$\eta'\pi$&$0.50\pm0.10$&0.16 &$<0.48$ &0.73\\
$\eta'\rho$&$< 0.52$ &0.05 &$<0.07$ & 0.12\\ \hline
\end{tabular}
\end{table}

	We gratefully acknowledge the effort of the CESR staff in 
providing us with
excellent luminosity and running conditions.
J.P.A., J.R.P., and I.P.J.S. thank                                           
the NYI program of the NSF, 
M.S. thanks the PFF program of the NSF,
G.E. thanks the Heisenberg Foundation, 
K.K.G., M.S., H.N.N., T.S., and H.Y. thank the
OJI program of DOE, 
J.R.P., K.H., M.S. and V.S. thank the A.P. Sloan Foundation,
R.W. thanks the 
Alexander von Humboldt Stiftung,
and M.S. thanks Research Corporation
for support.
This work was supported by the National Science Foundation, the
U.S. Department of Energy, and the Natural Sciences and Engineering Research 
Council of Canada.


\begin{thebibliography}{99}
\bibitem{Kamal2} Note that the approximate equality
of the two ratios is coincidental. See ~\cite{Kamal1}.
\end{thebibliography}
\end{document}